\documentclass[e-only,10pt,reqno]{ofj}

\usepackage{tensorNotation}
\usepackage{finiteVolumeNotation}

\usepackage{setspace}
\usepackage{lineno}
\usepackage{color}
\usepackage[dvipsnames,svgnames,x11names]{xcolor}
\usepackage[hang,flushmargin]{footmisc}
\usepackage{orcidlink}
\usepackage{cite}

\usepackage{todonotes}
\usepackage{xspace} 
\usepackage{caption}
\usepackage{subcaption}
\usepackage{makecell} 

\usepackage{chngcntr} 
\usepackage{orcidlink}
\usepackage{placeins}

\usepackage[colorlinks=true,linkcolor=black,anchorcolor=black,citecolor=black,filecolor=black,menucolor=black,runcolor=black,urlcolor=black]{hyperref}
\usepackage{cleveref} 

\newcommand{\authorcontributions}[1]{%
\vspace{6pt}\noindent{\fontsize{9}{11.2}\selectfont\textbf{Author Contributions:} {#1}\par}}

\let\paragraph\undefined

\newcommand{\OF}[0]{OpenFOAM\textsuperscript{\textregistered}\xspace}

\OpenFOAMversions{\OF v2412}

\ifdefined\review
\Repository{}
\else
\Repository{https://github.com/OpenFOAM-HPC-Challenge/OHC1}
\fi

\DOI{TBD} 

\ifdefined\review
\fi

\begin{document}


\title[\OF HPC Challenge (OHC-1)]{The First \OF HPC Challenge (OHC-1)}

\ifdefined\review
  \author{}
  \address{}
  \email{}
\else



\author[S. Lesnik]{Sergey Lesnik$^{1}$\orcidlink{0009-0007-5703-4438}}
\address{$^1$Wikki GmbH, Wernigerode, Germany}
\email{sergey.lesnik@wikki-gmbh.de}

\author[G. Olenik]{Gregor Olenik$^{2}$
\orcidlink{0000-0002-0128-3933}}
\address{$^2$Technical University of Munich~(TUM), Munich, Germany}
\email{gregor.olenik@tum.de}

\author[M. Wasserman]{Mark Wasserman$^{3}$\orcidlink{0009-0002-7327-0872}}
\address{$^3$Toga Networks - a Huawei Company, Haifa, Israel}
\fi


\begin{abstract}
  The first OpenFOAM HPC Challenge (OHC-1) was organised by the OpenFOAM HPC Technical Committee (HPCTC) to collect a snapshot of \OF's computational performance on contemporary production hardware and assess the impact of software optimisations on computational efficiency.
  Participants ran a common incompressible steady-state RANS case, the open-closed cooling DrivAer (occDrivAer) configuration, on prescribed meshes, submitting either with the reference setup (hardware track) or with modified solvers, decomposition strategies, or accelerator offloading (software track).

  In total, 237 valid data points were submitted by 12 contributors: 175 in the
  hardware track and 62 in the software track. The hardware track covered 25
  distinct CPU models across AMD, Intel, and ARM families, with runs spanning from
  single-node configurations up to 256 nodes (32\,768 CPU cores). Wall-clock times
  ranged from 7.8~minutes to 65.7~hours and reported energy-to-solution from 2.1
  to 236.9\,kWh. Analysis of the hardware track identified a Pareto front of
  optimal balance between time- and energy-to-solution, and revealed that
  on-package high-bandwidth memory (HBM) dominates single-node performance for
  next-generation CPUs.

  Software-track submissions achieved up to 28\% lower energy per iteration,
  17\% higher performance per node, and 72\% shorter time per
  iteration than the best hardware-track results, with full GPU ports and
  selective-memory optimisations leading the performance range. This manuscript
  describes the challenge organisation, the case setup and metrics, and presents
  the main findings from both tracks together with an outlook for future
  challenges.
\end{abstract}

\date{\today}

\dedicatory{}

\maketitle

\ifdefined\review
  \linenumbers
\else
\fi

\tableofcontents


\section{Introduction}

OpenFOAM is extensively utilised in both academia and industry for research and
design, with many applications requiring large computational resources. It is
therefore common to find OpenFOAM among the principal consumers of compute capacity on
High Performance Computing (HPC) clusters. To support future improvements in resource efficiency, the first
OpenFOAM HPC Challenge (OHC-1) was launched by the OpenFOAM HPC Technical
Committee (HPCTC). In this context, challenge problems are a commonly used tool to test hardware performance or to compare algorithmic approaches and
implementations. Well-known examples include the High Performance Linpack (HPL)
from the HPCC suite~\cite{dongarra2017hpc}, which underpins the TOP500 list, and
domain-specific challenges in areas such as automotive
CFD~\cite{autocfd2,autocfd3,autocfd4}. The term ``challenge'' is used here to
encourage community contributions aimed at improving computational efficiency
rather than only reporting baseline performance.

The goal of OHC-1 was to gather a snapshot of \OF's performance on current production
hardware and to enable comparison of hardware configurations with software
optimisations. 
Furthermore, the initiative aimed to:
\begin{itemize}
\item collect data for evaluating future algorithmic and implementation choices
  in \OF;
\item assess efficiency and scalability across hardware platforms;
\item compare performance of recent \OF variants and software optimisations;
\item propose unified metrics to guide development on next-generation systems;
\item promote a more relevant community benchmark case than the commonly used
  Lid-Driven Cavity 3D case 
\footnote{\url{https://develop.openfoam.com/committees/hpc/-/tree/develop/incompressible/icoFoam}}.
\end{itemize}
The results of OHC-1 were presented as a mini-symposium at the 20\textsuperscript{th} OpenFOAM Workshop in Vienna
on July~1, 2025. The challenge attracted submission of 237 data points contributed by 12
organisations from academia and industry.
The combined reported energy consumption of all submissions was approximately
4715\,kWh.

The manuscript is structured as follows.
\Cref{sec:setup} presents the case setup and mesh options.
\Cref{sec:rules} describes the organisation of the HPC Challenge, the metrics, and the validation criteria.
\Cref{sec:analysis} summarises the submitted data points and \Cref{sec:hardware_track,sec:software_track} present the main findings of the hardware and software tracks.
\Cref{sec:conclusion} concludes with a summary and outlook for future challenges.
\Cref{sec:repository} describes the public data repository where all submissions and analysis tools are available.

\section{Case Setup}\label{sec:setup}\label{sec:case-setup}

The HPCTC selected a single test case that had to be representative of typical
industrial usage, computationally demanding, and suitable as a surrogate for
common simulation workloads without relying on features that would exclude
alternative implementations. Based on these criteria, the case with the
external aerodynamic flow over a static DrivAer automotive model has been chosen. It had
already been studied in a similar form as case~2 in the AutoCFD2
workshop~\cite{autocfd2} and as case~2a in the AutoCFD3 and AutoCFD4
workshops~\cite{autocfd3,autocfd4}, so reference results were available to
validate submissions.

The challenge case is the open-closed cooling DrivAer (occDrivAer) configuration: external aerodynamic flow around a full car with closed cooling inlets and static wheels.
The geometry is based on the notchback variant of the Ford Open Cooling DrivAer~(OCDA), with a complex underbody and engine bay cooling channels.
For OHC-1 it was configured as a steady-state, incompressible RANS simulation using the SIMPLE algorithm with fixed inner iterations. Several meshes were pre-generated using snappyHexMesh with three resolutions: 65, 110 and 236 million cells, which were denoted as 65M, 110M and 236M, respectively.
The case is available in the HPCTC repository\footnote{\url{https://develop.openfoam.com/committees/hpc/-/tree/develop/incompressible/simpleFoam/occDrivAerStaticMesh}}.
The $k$--$\omega$ SST turbulence model was employed with pre-defined differencing schemes (\texttt{fvSchemes}).
Key case parameters are given in \Cref{tab:case_properties}.
The Reynolds number based on wheelbase $L_\text{ref}$ and inlet velocity is approximately $7.2\times 10^6$.

\begin{table}[!ht]
  \caption{Case properties for the occDrivAer challenge case.}
  \label{tab:case_properties}
  \begin{tabular}{lrc}
    \toprule
    Kinematic viscosity (Air) $\nu$ & $1.507 \cdot 10^{-5}$ & $\mathrm{m}^2/\mathrm{s}$ \\
    Inlet velocity $U$ & $38.889$ & $\mathrm{m}/\mathrm{s}$ \\
    Wheelbase $L_\text{ref}$ & 2.78618 & m \\
    Reynolds number Re & $7.1899\cdot 10^{6}$ & -- \\
    \bottomrule
  \end{tabular}
\end{table}

\begin{figure*}[!ht]
  \centering
  \begin{subfigure}{0.32\textwidth}
    \includegraphics[width=\textwidth]{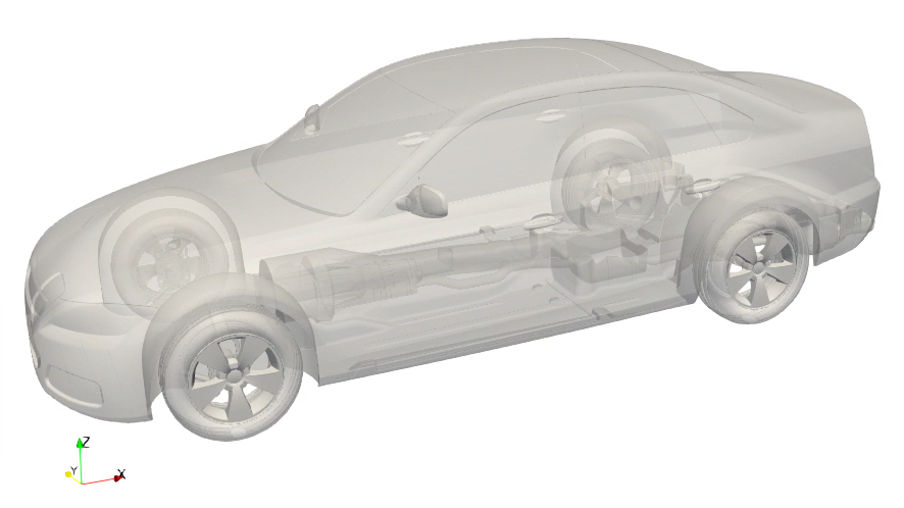}
    \caption{occDrivAer geometry.}
    \label{fig:occDrivAer}
  \end{subfigure}
  \hfill
  \begin{subfigure}{0.32\textwidth}
    \includegraphics[width=\textwidth]{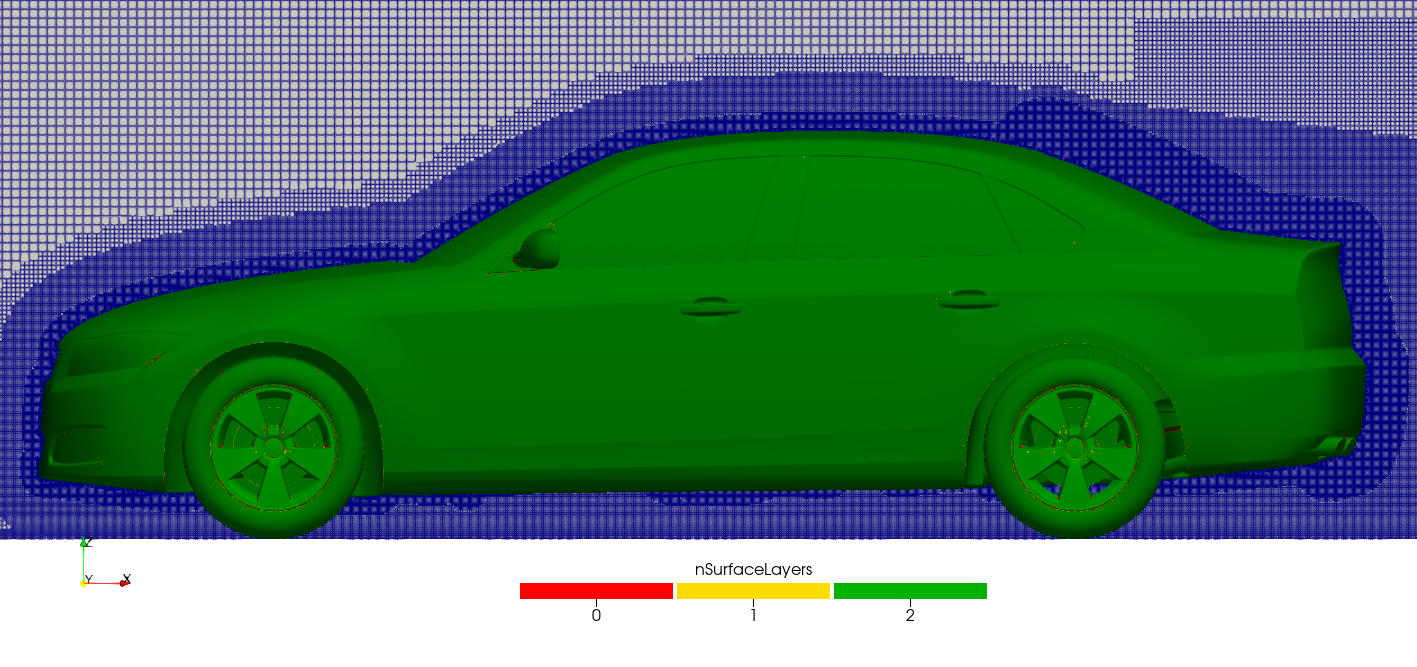}
    \caption{Mid-plane mesh.}
    \label{fig:mesh_mid}
  \end{subfigure}
  \hfill
  \begin{subfigure}{0.32\textwidth}
    \includegraphics[width=\textwidth]{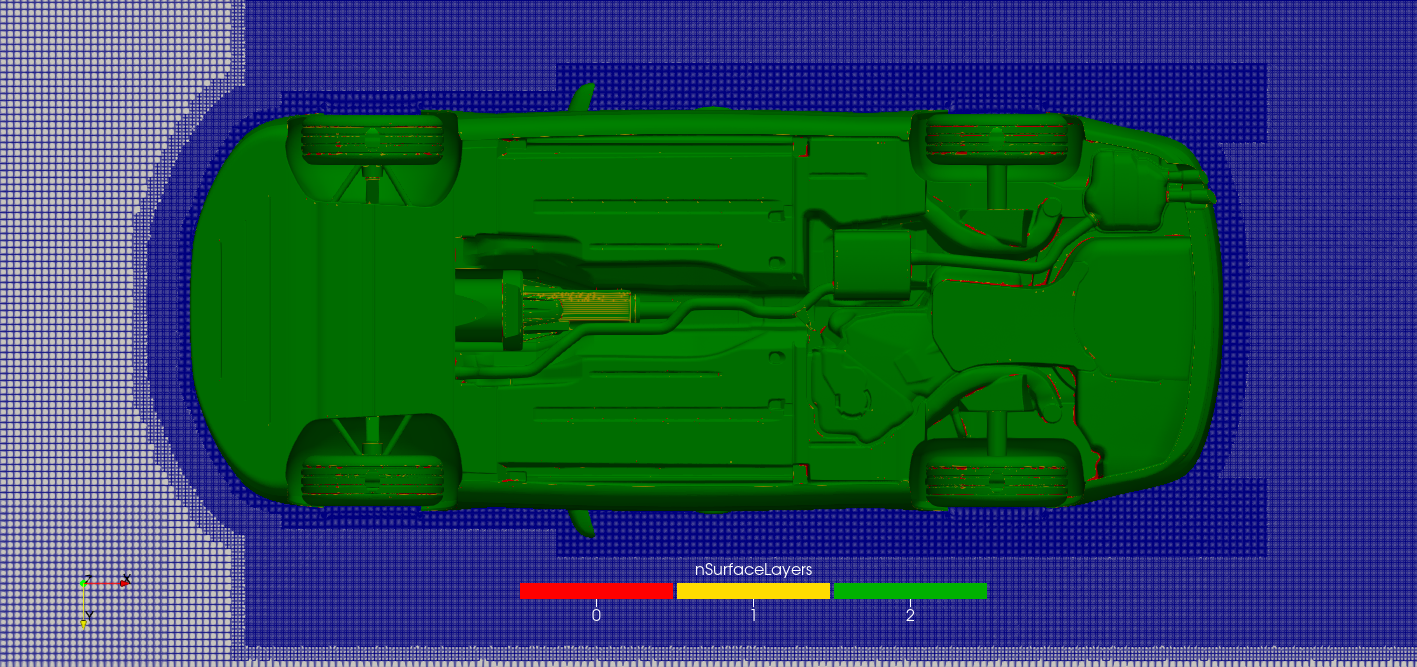}
    \caption{Underbody mesh (cut at axle).}
    \label{fig:mesh_underbody}
  \end{subfigure}
  \caption{Challenge case geometry and mesh.}
  \label{fig:case_geometry}
\end{figure*}

\Cref{fig:case_geometry} shows the geometry and representative mesh views: the full occDrivAer configuration, a mid-plane section, and an underbody cut.

\section{Challenge Structure and Procedure} \label{sec:rules}

\begin{figure*}[!ht]
  \centering
  \includegraphics[width=0.9\textwidth]{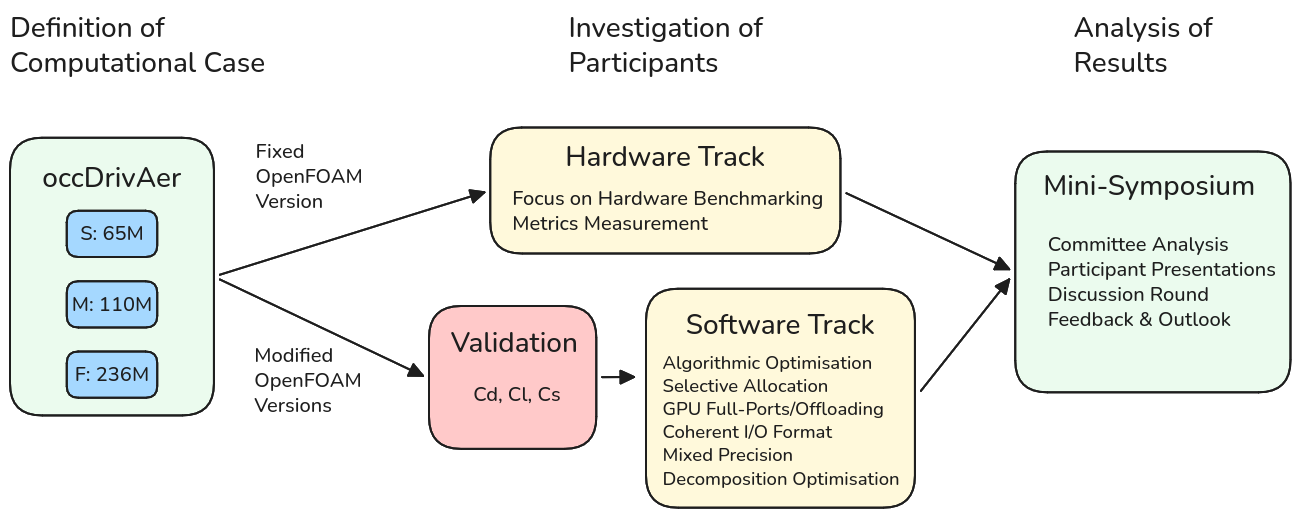}
  \caption{Overview of the OHC-1 procedure.}
  \label{fig:challenge_procedure}
\end{figure*}

After defining the computational scope of the challenges, the procedure was structured as illustrated in \cref{fig:challenge_procedure}. Data submissions were categorised into two tracks:
\begin{description}
\item[Hardware track.] Participants used the provided case setup and the official
  \OF v2412 release. Only the hardware, number of nodes and cores, and system software
  stack could vary.
\item[Software track.] Any \OF version and custom solvers or modifications were
  allowed, provided the mesh and physical model remained unchanged and solution
  The proposed topics for investigation in the software track included accelerators, pre-/post-processing (I/O), mixed-precision arithmetic, linear solvers, and renumbering/decomposition strategies. Submissions were accepted after the validation of computational results described in section~\ref{sec:validation}.
\end{description}

The submitted results were then analysed by the organising committee based on a set of predefined metrics. A comprehensive analysis of the hardware track and software track results was subsequently presented by the organisers in the form of a mini-symposium at the 20\textsuperscript{th} \OF Workshop. In addition, participants delivered in-depth presentations discussing their individual contributions and investigations.

\subsection{Metrics}\label{sec:metrics}

The following metrics were used to evaluate and compare the submissions in both tracks.

\begin{description}
\item[Time-to-solution (TTS)] The wall-clock time from the start of the main
  solver to completion of the prescribed number of iterations, excluding
  pre-processing and initialisation.
\item[Energy-to-solution (ETS)] The total energy consumed during the solver
  run.  Where direct power measurements were not available, energy was estimated
  from the Thermal Design Power~(TDP) of the CPUs/GPUs.
\item[Pre-processing wall-clock time] The wall-clock time for running the utilities that prepare the case for the run of the main solver application, e.g. decomposition, renumbering, initialisation with potentialFoam.
\item[Time per iteration] The mean wall-clock time of a single SIMPLE iteration.
\item[Energy per iteration] The mean energy consumed per SIMPLE iteration, derived from ETS divided by the number of iterations.
\item[FVOPS (Finite-Volume Operations Per Second)] A throughput metric proposed in~\cite{Galeazzo2024a} defined as the number of mesh cells multiplied by the number of iterations, divided by the wall-clock time:
  \begin{equation}
    \text{FVOPS} = \frac{N_\text{cells} \times N_\text{iter}}{t_\text{WC}}\,.
    \label{eq:fvops}
  \end{equation}
  FVOPS is often reported per node to enable comparison across different node counts.
\item[FVOPS per energy] The ratio of FVOPS to energy, giving an
  energy-efficiency throughput metric.
\end{description}
During the data collection and analysis process, several corrections and
assumptions were necessary:
\begin{itemize}
\item The number of cores was corrected to represent \emph{physical} cores;
  hyper-threading was disregarded.
\item If not reported by the participants, some values were added based on publicly available data
  (e.g.\ TDP according to the CPU model).
\item The energy analysis is \emph{mostly based on theoretical TDP values} rather
  than direct power measurements, which introduces uncertainty.  Only a small
  number of submissions provided energy data from total cluster measurements.
\item The focus of this work lies on providing an overview of the complete data
  set; analysis of individual simulations or submissions are outside of the scope. When available references to further publications for in-depth analysis is provided.
\end{itemize}

\subsection{Submission Validation} \label{sec:validation}
Since Hardware-track submissions used the prescribed setup. Their correctness was
inherent in the use of the reference case. In contrast, Software-track submissions required
additional validation to ensure that solution quality was preserved despite any modifications in the software, case decomposition, or employed algorithms. Participants were required to extract force
coefficient convergence histories at run time and analyse the mean and variance of the aerodynamic coefficients ($C_d$, $C_l$, $C_s$) using the \texttt{meanCalc}\footnote{\url{meancalc.upstream-cfd.com/}}
utility. The validity criterion was
\begin{equation}
  2\,\sigma(\mu) < 0.0015\,,
  \label{eq:validation}
\end{equation}
where $\sigma(\mu)$ denotes the standard error of the running mean $\mu$ of the respective force coefficient. Reference values obtained with standard \OF on the
fine mesh were
$C_d = 0.262$,
$C_l = 0.0787$, and
$C_s = 0.0116$.
The data analysis team double-checked and cross-compared reported values with the submitted time series data.

\begin{figure*}[!ht]
  \centering
  \includegraphics[width=0.95\textwidth]{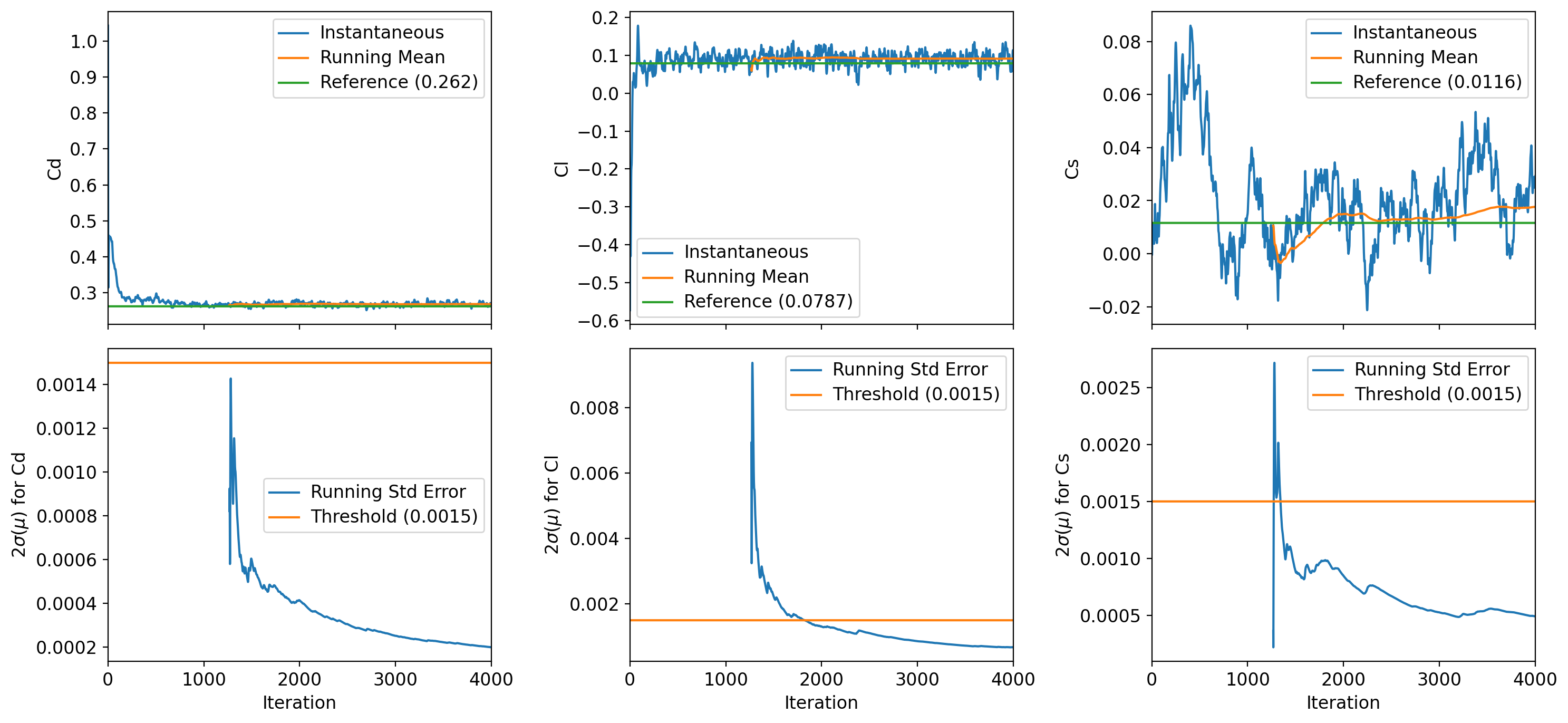}
  \caption{Example meanCalc validation for a software-track submission. Top row: instantaneous force coefficients (blue) with running mean from iteration~1265 (orange) and reference value (green). Bottom row: convergence of $2\,\sigma(\mu)$ (blue) towards the acceptance threshold of~0.0015 (orange).}
  \label{fig:meancalc_example}
\end{figure*}

\Cref{fig:meancalc_example} illustrates the validation process for an example software-track submission.
The top row shows the instantaneous force coefficients together with the running mean computed from iteration~1265 onwards.
In this case, the iterations before iteration~1265 were identified by meanCalc as initialisation transients and thus were not taken into account for the evaluation.
The bottom row shows the running double standard error of the sample mean $2\,\sigma(\mu)$, which must fall below the threshold of~0.0015 for each coefficient.
For this submission the final values are all within the acceptance criterion.
The converged running-mean coefficients ($C_d = 0.2682$, $C_l = 0.0911$, $C_s = 0.0177$) are in reasonable agreement with the reference values, confirming that the software modifications preserved solution accuracy.

\section{Data Analysis} \label{sec:analysis}


OHC-1 received contributions from 12 organisations spanning academia, research
centres, and industry:
Wikki~GmbH,
University College Dublin~(UCD),
CFD FEA Service,
CINECA,
Huawei,
Universit\"at der Bundeswehr M\"unchen,
Federal Waterways Engineering and Research Institute~(BAW),
University of Minho,
Technical University of Munich~(TUM),
United Kingdom Atomic Energy Authority~(UKAEA),
Engys, and
E4 Computer Engineering.

A total of 237 valid data points were submitted: 175 in the hardware track and 62 in the software track.
Runs ranged from single-node configurations up to 256 compute nodes (32\,768 CPU cores).
Wall-clock time to solution ranged from 7.8~minutes to 65.7~hours -- a factor of about 560.
Reported energy-to-solution (when provided) ranged from 2.1 to 236.9\,kWh -- a factor of about 110, resulting in a combined reported consumption of approximately 4715\,kWh across all submissions.
\Cref{tab:overview-sub-data} provides a summary of the key figures.

\begin{table}[!ht]
  \caption{Overview of submitted data points (OHC-1).}
  \label{tab:overview-sub-data}
  \begin{tabular}{lc}
    \toprule
    Submitted data points (total / hardware / software) & 237 / 175 / 62 \\
    CPU models (distinct) & 25 \\
    GPU models (distinct) & 3 \\
    Vendor split (AMD / Intel / ARM) & 106 / 80 / 50 \\
    Max compute nodes & 256 \\
    CPU cores (min / max) & 1 / 32\,768 \\
    Wall-clock time to completion (min / max) & 7.8~min / 65.7~h \\
    Energy to solution (min / max), when reported & 2.1 / 236.9~kWh \\
    Total reported energy consumption & $\approx 4715$~kWh \\
    \bottomrule
  \end{tabular}
\end{table}

The analysis is organised as follows. The hardware track (\Cref{sec:hardware_track})
examines the energy-time trade-off and identifies the Pareto front, analyses
strong scaling behaviour, isolates single-node compute performance from
interconnect effects, and discusses the impact of on-package high-bandwidth memory.
The software track (\Cref{sec:software_track}) is discussed by optimisation
category and compared to the hardware track in terms of FVOPS per node and energy
per iteration.

\subsection{Hardware Track} \label{sec:hardware_track} \label{sec:res_hardware_track}

\begin{figure}
     \centering
     \begin{subfigure}[b][5cm][]{0.30\textwidth}
         \centering
        \includegraphics[height=5cm]{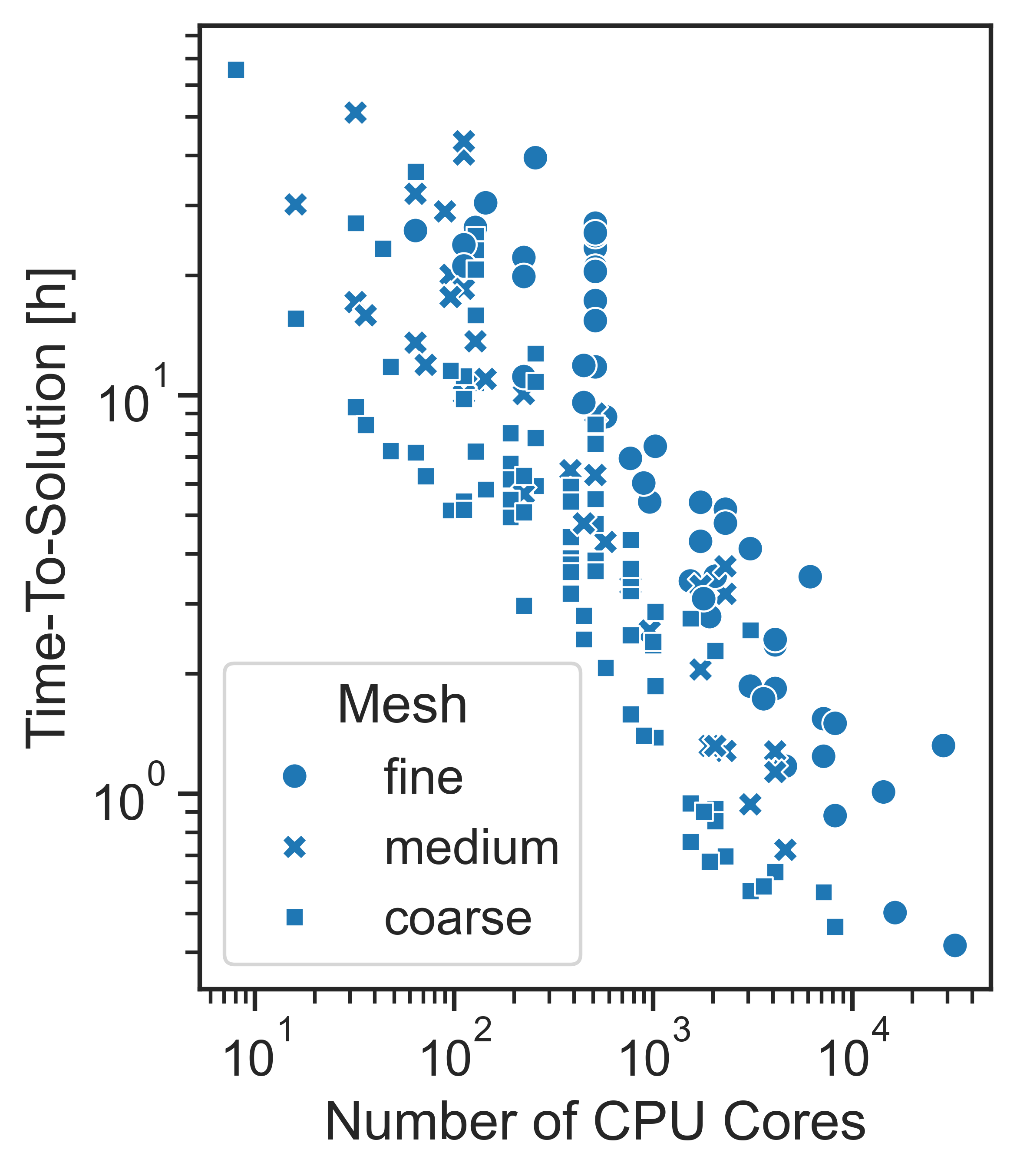}
        \caption{\label{fig:overview_hardware}}
     \end{subfigure}
     \hfill
     \begin{subfigure}[b][5cm][]{0.14\textwidth}
         \centering
         \includegraphics[height=5cm]{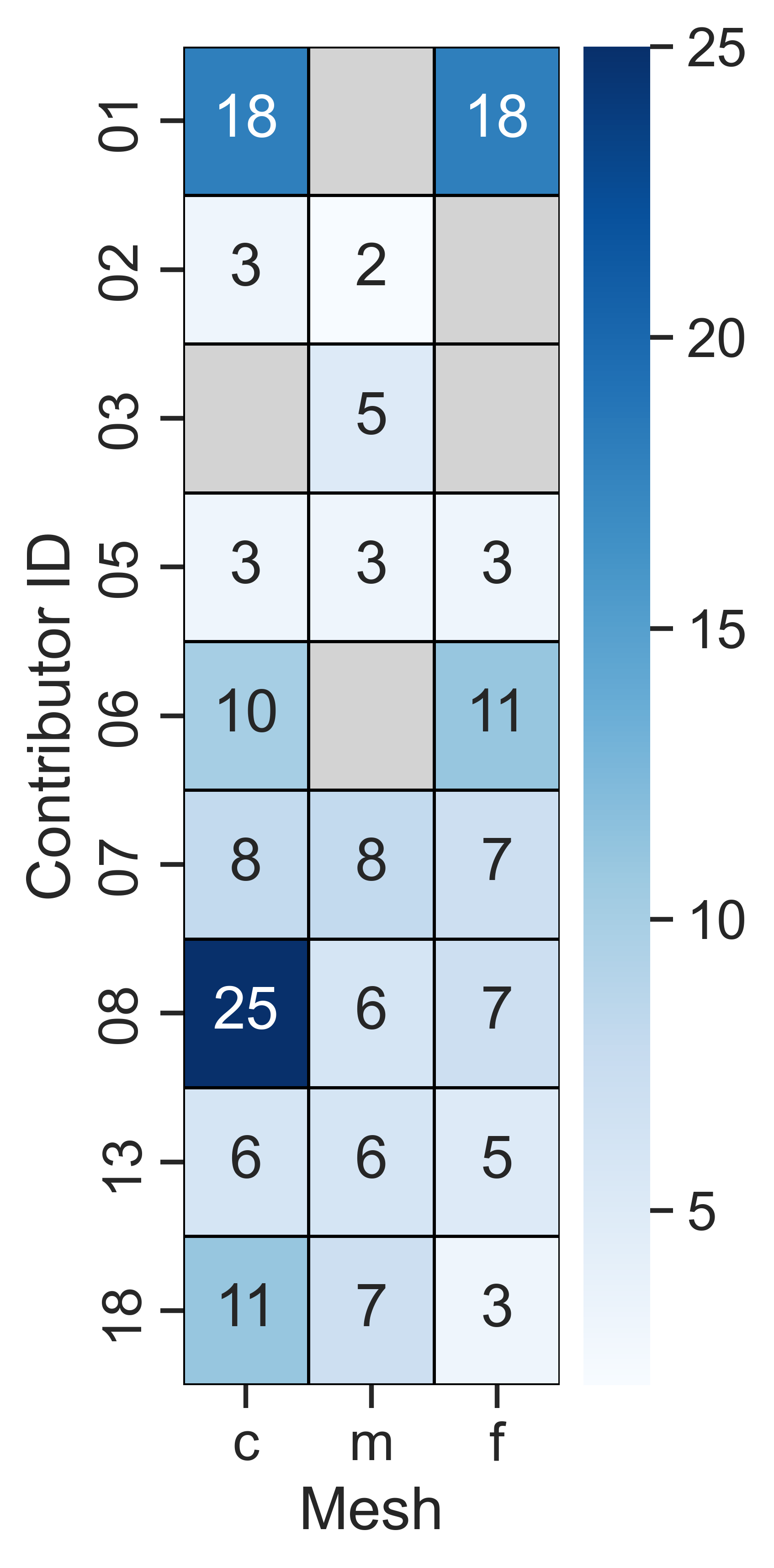}
         \caption{\label{fig:overview_mesh_heatmap}}
     \end{subfigure}
     \hfill
     \begin{subfigure}[b][5cm][]{0.50\textwidth}
         \centering
\includegraphics[height=5cm]{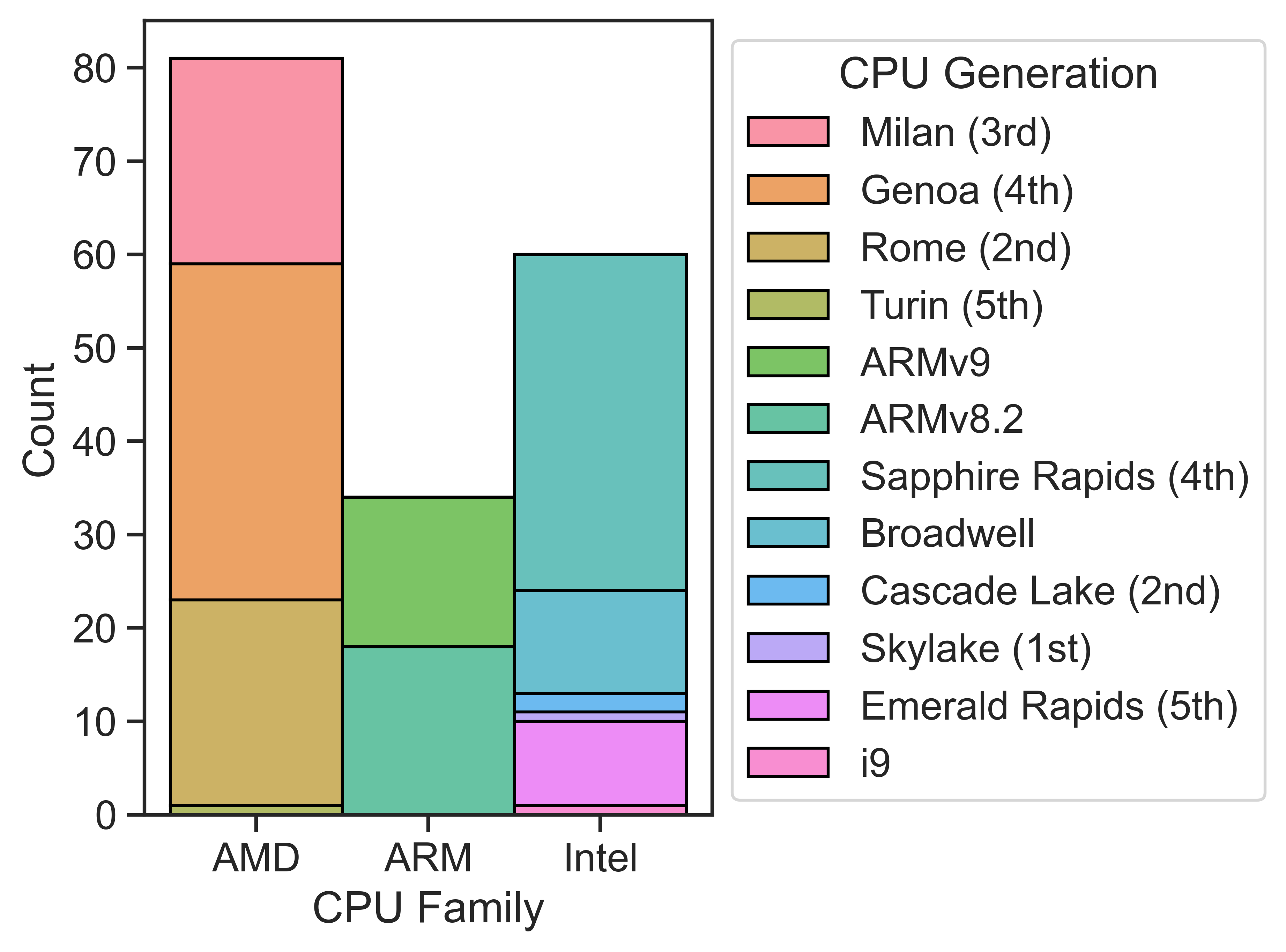}             \caption{\label{fig:overview_hardware_cpu}}
     \end{subfigure}
    \caption{Overview of hardware-track submissions, (a) Time-to-solution over number of utilized CPU cores, (b) number of submissions based on mesh type and contributor ID, (c) number of submissions based on CPU vendor and CPU Generation.\label{fig:overview_hardware_track}}
\end{figure}

Hardware-track submissions used the reference case and an official \OF release, whereby only hardware, node count, and system software could vary.
In total, 175 data points were submitted. \Cref{fig:overview_hardware_track} gives an overview of the submissions, showing the reported time-to-solution over a broad set of employed CPUs and the different mesh sizes. As shown in \cref{fig:overview_mesh_heatmap} all three mesh sizes had numerous submissions, with some participants, e.g.\ 5, 7, and 13 submitting an equal amount of data points for all meshes. Some participants had a stronger focus on either coarse or fine mesh investigations, e.g.\ participants 8 and 6. In total, most data points where submitted for the coarse mesh and the least number for the medium, likely because participants either focused on the single node performance or the extreme scaling results. Furthermore the submitted data in the hardware track covers 20 distinct CPU models across three major vendor families: AMD (81 submissions), ARM~(34), and Intel~(60). \Cref{fig:overview_hardware_cpu} shows a breakdown of the number of  submissions into CPU vendors and individual CPU generations.

\subsubsection{Energy--time trade-off and Pareto front}

\begin{figure}[!ht]
  \centering
  \includegraphics[width=0.95\textwidth]{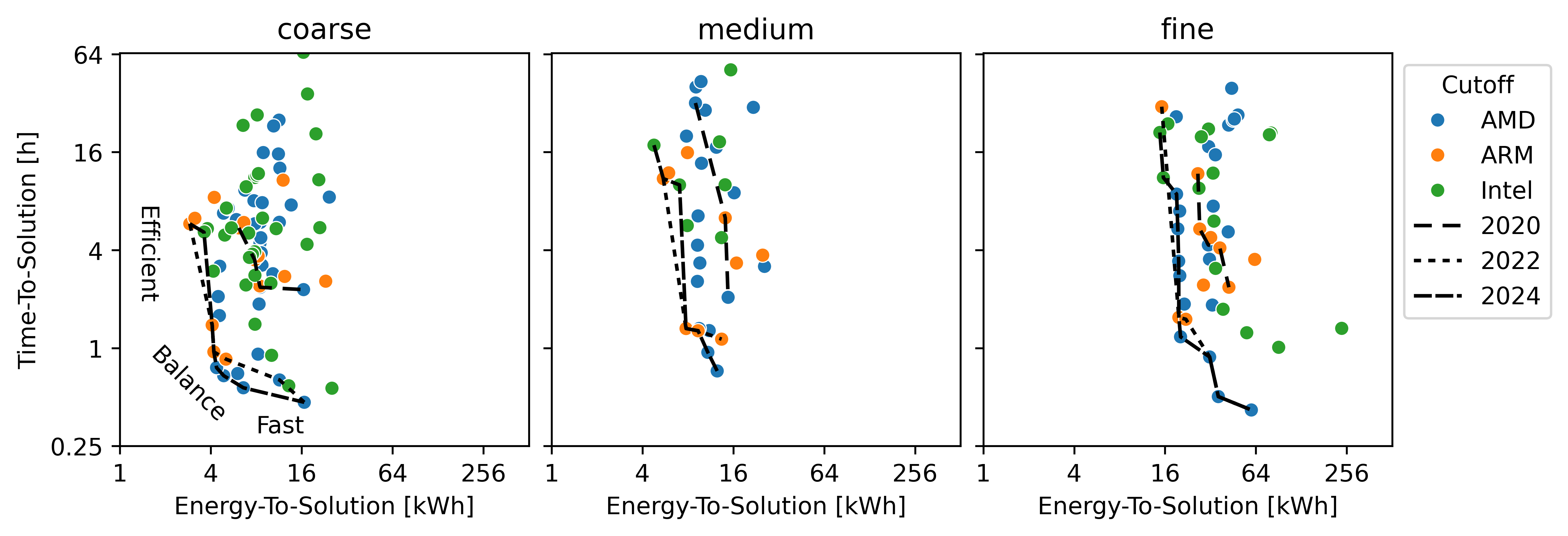}
  \caption{Energy-to-solution vs.\ time-to-solution for hardware-track runs,
    faceted by mesh size. Operating regimes and the Pareto front of optimal
    energy--time balance are indicated.}
  \label{fig:TTS_ETS_MESH_singleNode}
\end{figure}

A central question in HPC workloads is the trade-off between time-to-solution (TTS) and energy-to-solution (ETS).
\Cref{fig:TTS_ETS_MESH_singleNode} presents such a trade-off for all hardware-track runs, faceted by mesh size. Pareto fronts are shown as black lines indicating the optimal balance between TTS and ETS achieved with hardware released prior to that year.  From the pareto fronts, three distinct operating regimes are visible:
\begin{enumerate}
\item \emph{Eficient}: runs that minimise energy consumption at the cost of
  longer execution times (typically single-node or small-scale runs on
  energy-efficient hardware).
\item \emph{Fast}: runs that minimise wall-clock time by using many nodes, at
  the expense of higher total energy.
\item \emph{Balance}: runs that achieve a practical compromise between
  energy and time.
\end{enumerate}
%
The shapes of the Pareto fronts for lower values of consumed energy demonstrate that relatively large improvements (lower values) of TTS can be achieved with low increases in ETS.
On the other hand, for high ETS the Pareto front confirms that increasing parallelism yields diminishing time savings while incurring substantially higher energy costs, consistent with strong-scaling saturation effects discussed below.
For each mesh, the three families occupy overlapping but distinct regions of
the plot. ARM submissions tend to cluster in the lower-energy region, reflecting
recent many-core, high-bandwidth-memory designs. AMD and Intel submissions span a
wider range of node counts and correspondingly wider ranges of both time and
energy. Across all three meshes, the same qualitative trend holds: runs using
more nodes reduce time-to-solution but shift upward and to the right in energy
consumption.
\subsubsection{Strong-scaling behaviour}\label{sec:hw_scaling}

\begin{figure*}[!ht]
  \centering
  \includegraphics[width=0.9\textwidth]{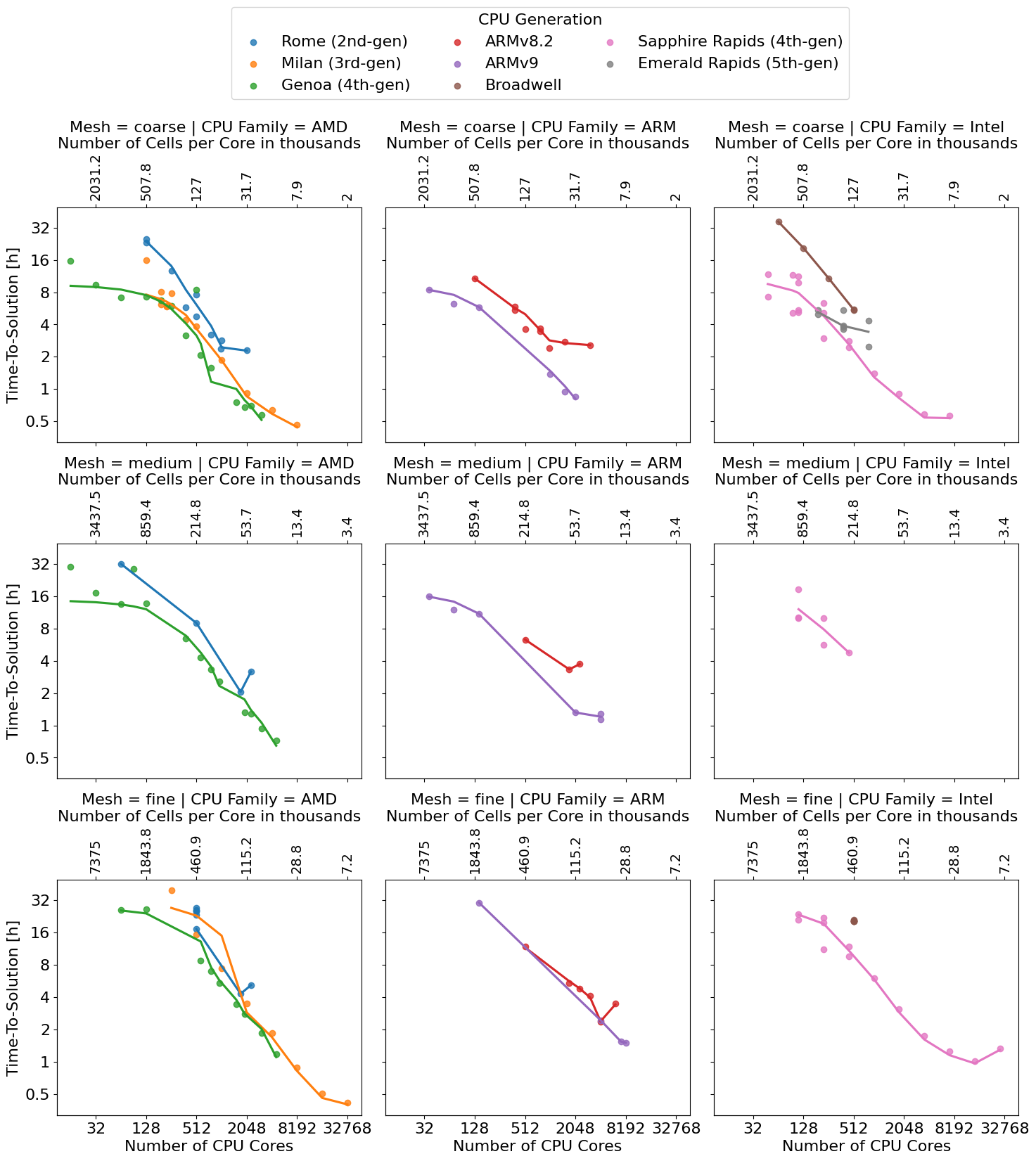}
  \caption{
  Strong scaling: time-to-solution vs.\ number of CPU cores for the coarse (top), medium (centre), and fine (bottom) meshes with a segmentation based on CPU family: AMD (left), ARM (centre), and Intel (right).
  Curves represent fits to the corresponding data points using a local regression model.}
  \label{fig:strong_scaling}
\end{figure*}

Strong scaling, which represents running a fixed problem size on an increasing number of CPU cores, reveals two distinct performance-limiting regimes:
\begin{itemize}
\item \emph{Single-node scaling limit:} Within a single node, performance is
  bounded by memory bandwidth.  As the number of cores per node increases,
  contention for shared memory buses limits the per-core throughput. This effect
  is especially pronounced for many-core architectures.
\item \emph{Multi-node scaling limit:} Beyond a single node, inter-node
  communication dominates the execution time. Since \OF relies exclusively on MPI
  for parallelism, the communication overhead grows with the surface-to-volume
  ratio of the domain decomposition. A practical scalability limit of
  approximately 10\,000~cells per core was observed across submissions. Below
  this threshold, communication costs exceed computational work. Depending on the CPU architecture, this limit may be larger.
\end{itemize}

\Cref{fig:strong_scaling} shows detailed strong-scaling behaviour across three mesh resolutions (coarse, medium, fine), three CPU families (AMD, ARM, Intel), and processor generations within each family.
Time-to-solution is plotted against core count, with the corresponding cells-per-core indicated along the top axis.
The data is represented by means of a scatter plot with the associated curves that are fitted using local regression.
Only data with more than two points per mesh and CPU generation is included in the plot to enable meaningful fitting.
Across all architectures and mesh sizes, time-to-solution decreases with increasing core count, confirming expected strong scaling behaviour.
However, none of the configurations exhibit ideal linear scaling over the full range.
In some cases, the slope of the curves inverts at higher core counts, indicating reduced parallel efficiency as the workload per core diminishes and communication overhead becomes more significant.
This is typically observed in the range of 10 to 50 thousand cells per core.
Within each CPU family, newer generations generally achieve lower absolute runtimes at comparable core counts. An alternative illustration of the scaling behaviour is presented in the appendix, \cref{fig:scaling_count_cpu}.

\subsubsection{Parallel efficiency}\label{sec:hw_parallel_eff}

\begin{figure*}[!ht]
  \centering
  \includegraphics[width=0.9\textwidth]{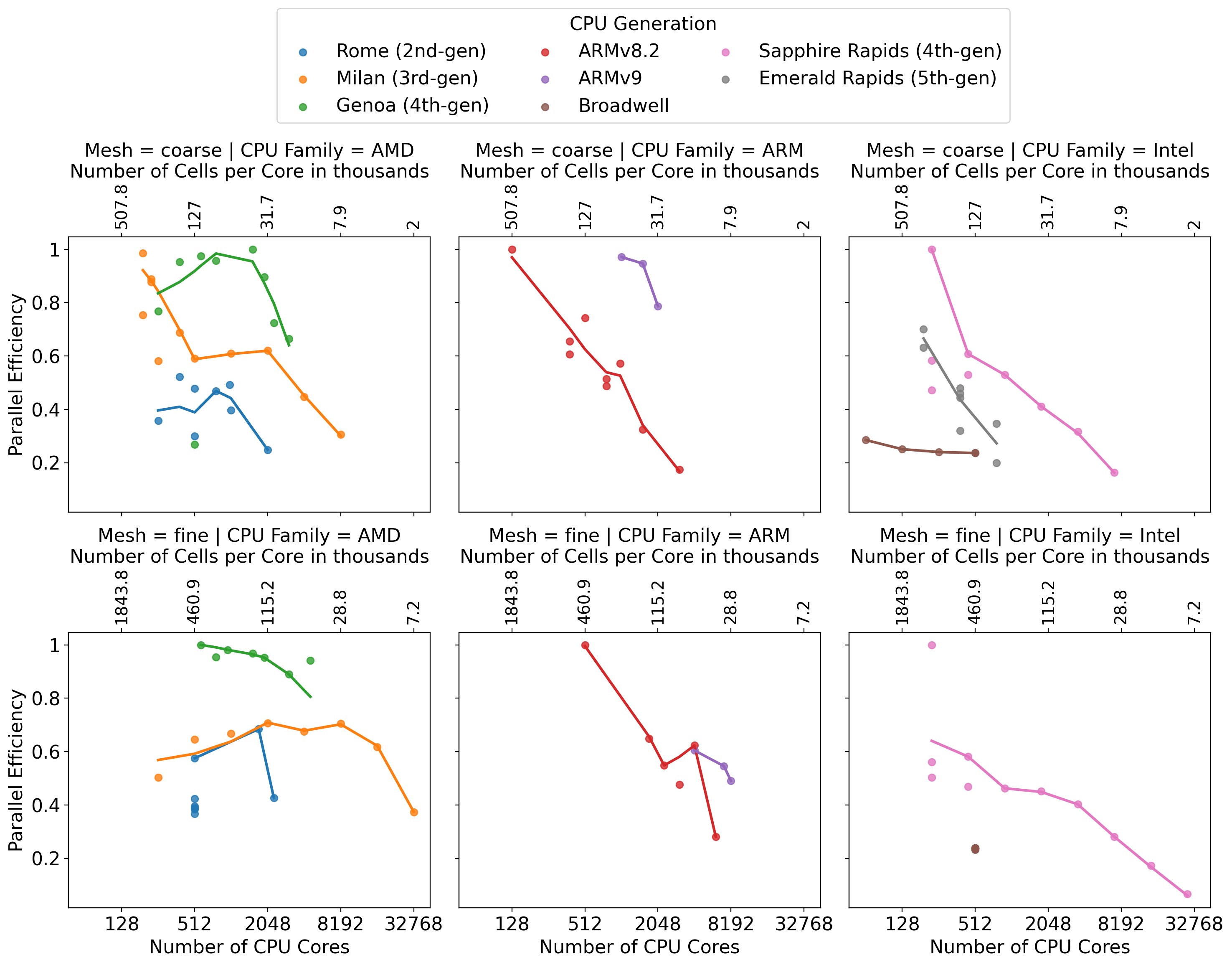}
  \caption{
   Parallel efficiency vs.\ number of CPU cores for the coarse (top) and fine (bottom) meshes with a segmentation based on CPU family: AMD (left), ARM (centre), and Intel (right).
   Curves represent fits to the corresponding data points using a local regression model.
    }
  \label{fig:parallel_efficiency}
\end{figure*}

To quantify how effectively additional cores reduce time-to-solution, we compute
a normalised parallel efficiency metric. For each CPU sub-model with
multi-node data, the efficiency is defined as the ratio of the minimum
core-weighted time (i.e.\ the most efficient single run) to the core-weighted
time of each run:
\begin{equation}
  \eta = \frac{\min_{i}(T_i \cdot N_i)}{T \cdot N}\,,
  \label{eq:parallel_eff}
\end{equation}
where $T$ is the time-to-solution, $N$ is the number of CPU cores, and the
minimum is taken over all multi-node runs for that sub-model.

\Cref{fig:parallel_efficiency} shows the parallel efficiency for the coarse and fine meshes categorised by CPU family.
The medium mesh was excluded from the analysis because of the insufficient amount of data points with multi-node runs.
The values are normalised according to the most efficient run from the corresponding mesh and CPU family.
Efficiency drops below unity as the core count increases, indicating communication overhead.
The rate of decline depends on CPU generation and vendor family: architectures with higher per-core memory bandwidth and more capable interconnects maintain higher efficiency at larger core counts.
Some of the CPU generations demonstrate higher parallel efficiency with the increasing number of cores indicating super-linear scaling, which is a known pattern for modern computer architectures~\cite{Galeazzo2024a}.

\subsubsection{High Bandwidth Memory and Last-Level Cache effect}\label{sec:hw_single_node}

\begin{figure*}[!ht]
  \centering
  \includegraphics[width=0.95\textwidth]{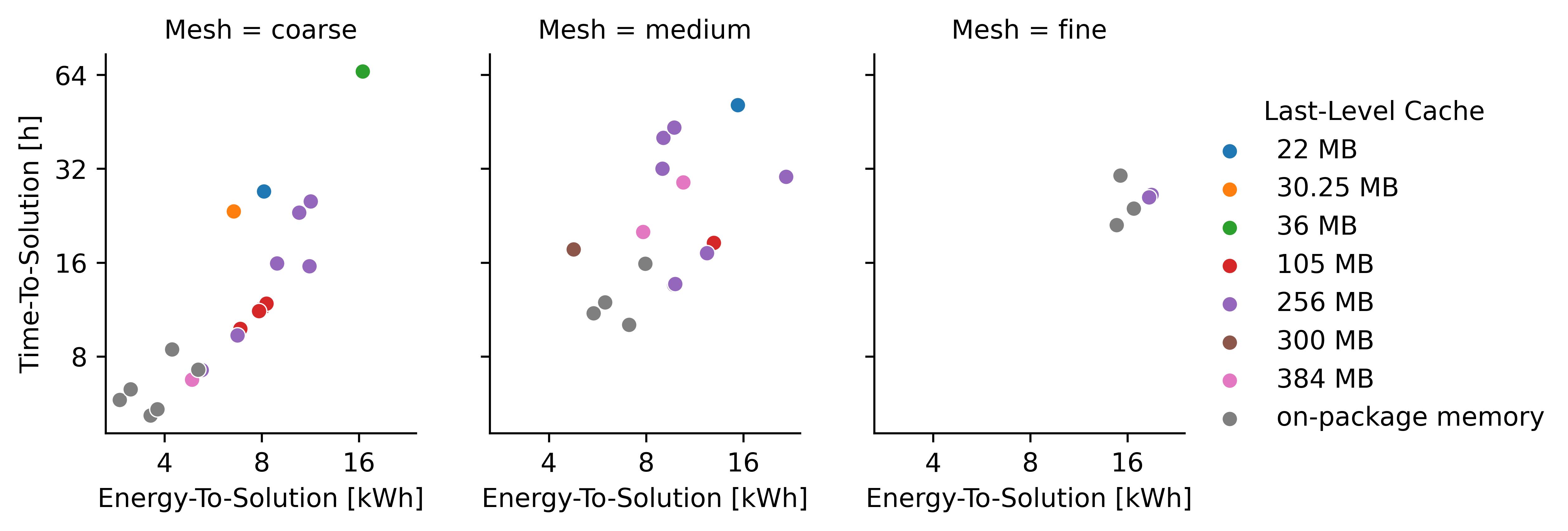}
  \caption{Single-node runs: time-to-solution vs.\ energy-to-solution, faceted by mesh size and coloured by last-level cache category.}
  \label{fig:singlenode_LLC}
\end{figure*}

Hardware vendors address the memory-bound bottleneck of the applications such as \OF by providing chip architectures with on-chip memory, e.g.\ High Bandwidth Memory (HBM) or large Last-Level Cache (LLC).
The following analysis provides insight on the impact of these approaches on the performance.
First, single-node runs are analysed to separate compute performance from interconnect effects.
\Cref{fig:singlenode_LLC} shows the single-node energy--time trade-off with data points coloured by last-level cache category.
For the coarse and medium meshes, submissions with larger HBM configurations consistently occupy the lower-left (faster, more efficient) region of the plot, demonstrating that on-package memory dominates single-node performance in terms of speed and efficiency.
This is consistent with the memory-bound nature of sparse iterative solvers used in \OF: the working set of \OF's field operations and linear-algebra kernels benefits directly from higher memory bandwidth.
In the case of the fine mesh, there were not enough submissions of single-node runs for an analysis.
The runs performed on the hardware with large last-level caches do not show a particular trend.
The amount of data for computing the case on a single node is orders of magnitude larger than the largest cache sizes.
Therefore, the computation is bound mainly by the bandwidth of the main memory.
This is different for the on-package memory chips, where the memory size is comparable of the problem size.

\begin{figure*}[!ht]
  \centering
  \includegraphics[width=0.95\textwidth]{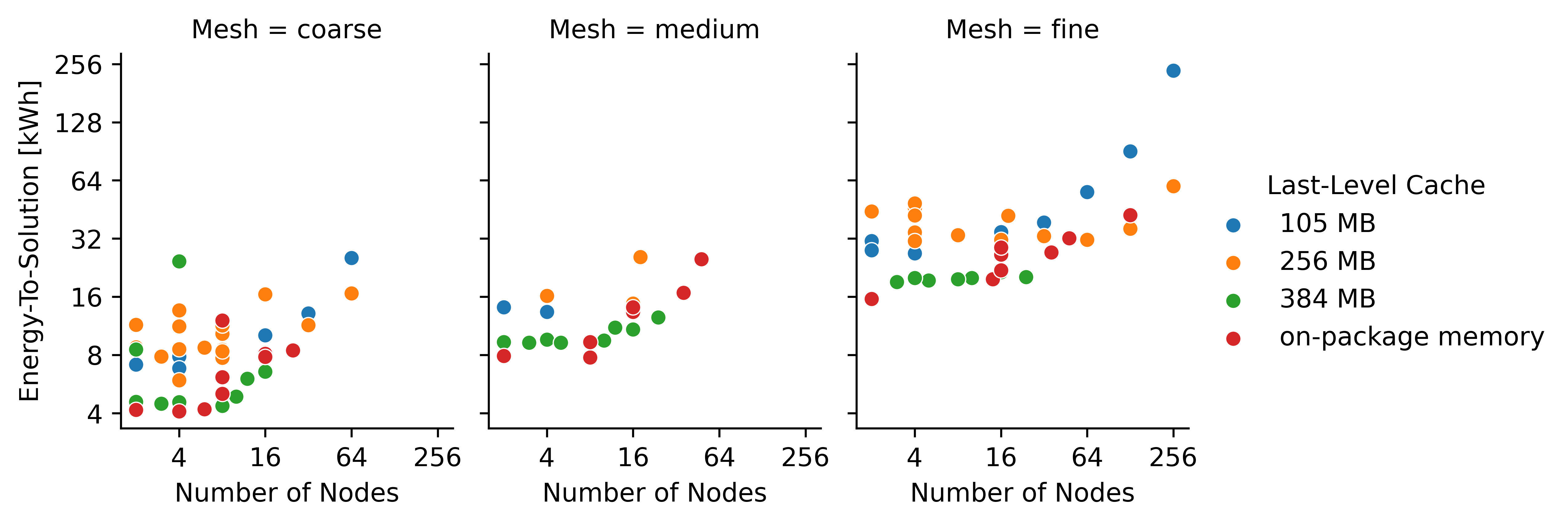}
  \caption{Multi-node runs: energy-to-solution vs.\ number of nodes, faceted by mesh size and coloured by last-level cache category.}
  \label{fig:ETS_nodes_facetGrid_mesh_LLC}
\end{figure*}

However, with the higher number of nodes the data load per CPU decreases.
An insight of the LLC size impact on ETS for such runs is provided in \cref{fig:ETS_nodes_facetGrid_mesh_LLC}, which presents data points only for the computations on more than one node.
Additionally, only three cache sizes together with the on-package memory cases were selected for a better overview.
The diagrams may be divided into two areas at around 8, 16, 32 nodes for the coarse, medium and fine mesh, respectively.
In the area of the lower node count, the ETS for particular LLC sizes tend to stay at the same level for the increasing number of nodes.
According to the observations from~\cite{Galeazzo2024a}, here, the growing communication overhead seems to be balanced by higher performance due to lower data amount per CPU.
In the area of the higher node count, this balancing does not seem to hold and the energy efficiency starts to deteriorate.
The runs with the 384\ MB LLC are more energy efficient than other large LLC computation.
This cannot be attributed to the cache size alone since the larger LLC belong to the newer hardware generations that may bring other architectural improvements.

\subsection{Software Track} \label{sec:software_track} \label{sec:res_software_track}

The software track allowed modifications to the default \OF setup to study for example the
impact of solver changes, decomposition strategies, or hardware-specific
optimisations. While hardware was unconstrained, the mesh and physical modelling
(spatial and temporal discretisation, turbulence model) had to remain
unchanged. Contributions from five organisations (Engys, CINECA, Huawei,
Wikki~GmbH, and KIT/TUM) fell into the following broad categories:
\begin{itemize}
\item Full GPU ports \cite{popovac_2025, bna2025spuma} (Engys, CINECA)
\item Offloading of linear solvers to GPUs via OGL \cite{Olenik2025Towards, olenik2025investigating} (KIT/TUM)
\item Decomposition optimisations \cite{popovac_2025} (Wikki, KIT/TUM)
\item Mixed precision \cite{Delorme2024} (Huawei)
\item Selective memory allocation \cite{popovac_2025}  (Huawei)
\item Coherent I/O format \cite{weiss2024coherent} (Wikki)
\end{itemize}
In the following only a high level overview of the impact of the optimisation techniques is presented, a full discussion of the individual techniques is outside the scope of this work and the interested reader is referred to individual publications and conference contributions in \cite{popovac_2025}.

\Cref{tab:comp-soft-hardw} summarises selected
metrics comparing the best software-track and hardware-track results.
The software track represented approximately 26\% of total submissions.

\begin{table}[!ht]
  \caption{Comparison of key software-track vs.\ hardware-track metrics.}
  \label{tab:comp-soft-hardw}
  \begin{tabular}{lccc}
    \toprule
    \thead{Metric} & \thead{Software Track} & \thead{Hardware Track} & \thead{Change w.r.t Hardware Track} \\
    \midrule
    Submitted results & 62 & 175 & 26\% of total \\
    Minimum energy per iteration [J] & 1867 & 2613 & $-28$\% \\
    Maximum performance [MFVOPS] & 737 & 629 & $+17$\% \\
    \bottomrule
  \end{tabular}
\end{table}

\subsubsection{Distribution of submissions}
\begin{figure}[!ht]
  \centering
    \begin{subfigure}[b][5cm][]{0.20\textwidth}
    \includegraphics[height=4.5cm]{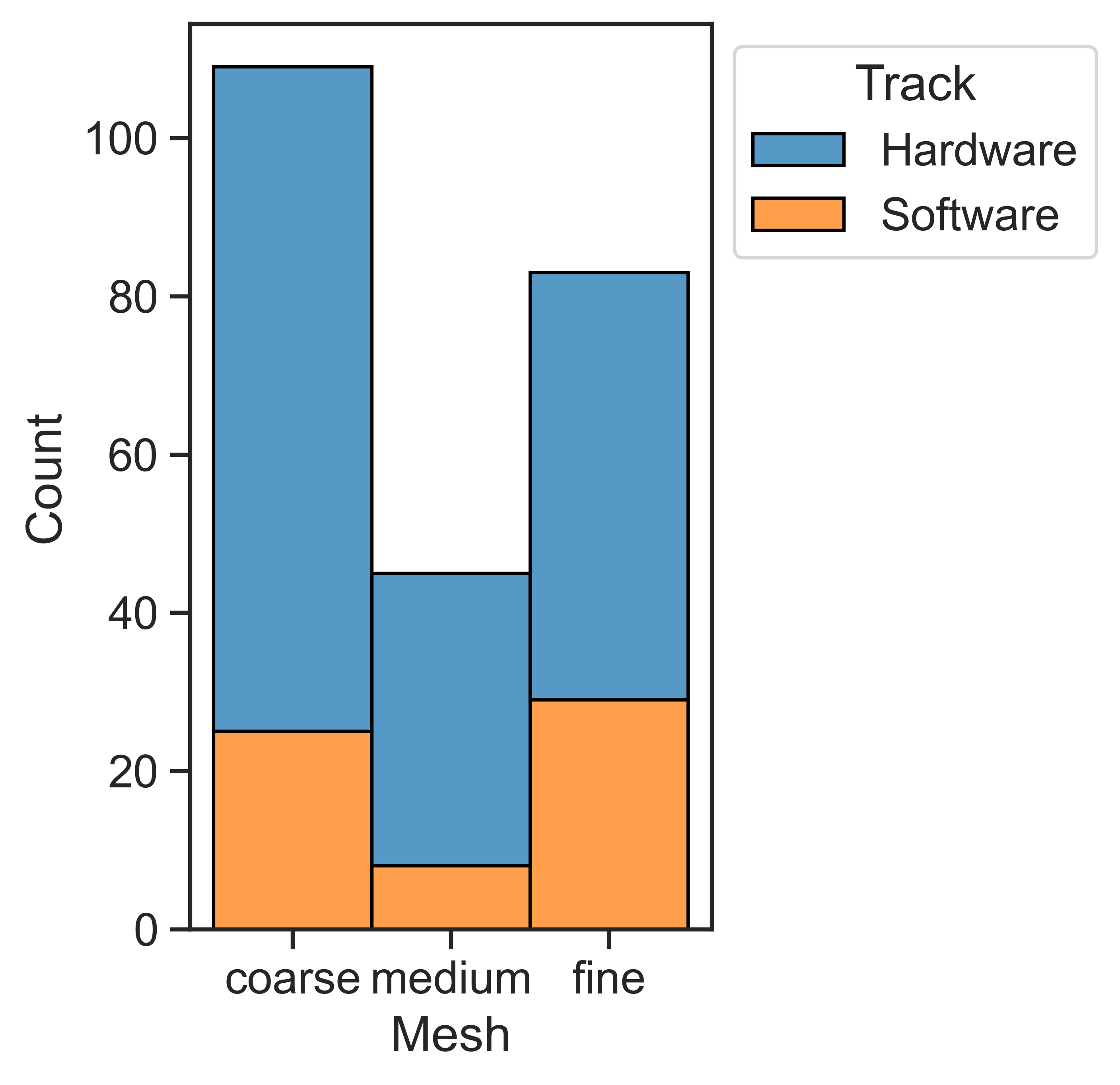}
    \caption{\label{fig:relation_hardware_vs_software_track_mesh}}
  \end{subfigure}
  \hfill
    \begin{subfigure}[b][5cm][]{0.70\textwidth}
    \includegraphics[height=4.5cm]{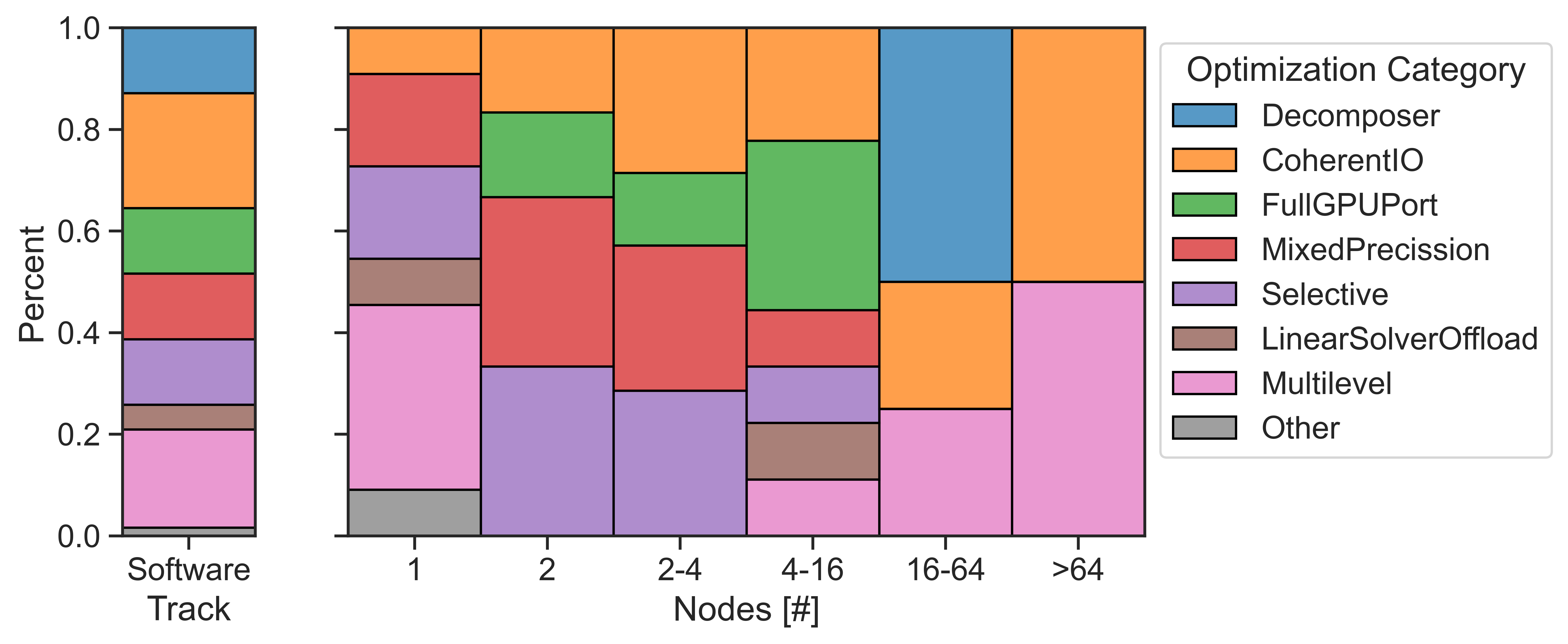}
    \caption{}
  \end{subfigure}
  \caption{Distribution of hardware and software track contributions by mesh type (a) and distribution of software optimisation category contributions over number of nodes (b).
  \label{fig:relation_hardware_vs_software_track}}
\end{figure}

\Cref{fig:relation_hardware_vs_software_track} illustrates the distribution of software-track submissions. \Cref{fig:relation_hardware_vs_software_track_mesh} shows the number of hardware- and software-track submissions across different mesh sizes.
Most software-track data points correspond to the fine mesh, both in absolute numbers and relative to the hardware-track submissions.
The breakdown of optimisation categories, shown in \Cref{fig:relation_hardware_vs_software_track}, indicates that the majority of submissions focused on coherent I/O and multilevel decomposition. These are followed by selective memory allocation, mixed-precision solvers, custom decomposition techniques, and full GPU ports.
An analysis of the distribution of the optimisation categories across different node counts reveals that several optimisation approaches, such as linear solver offloading, full GPU ports, mixed-precision methods, and selective memory allocation, were primarily investigated at smaller scales, up to 16 nodes. In contrast, multilevel decomposition and coherent I/O were also explored at larger scales, exceeding 64 nodes. Custom communication-optimized decomposition strategies were studied in the intermediate range, between 16 and 64 nodes. In addition to the CPU models evaluated in the hardware track, nine submissions utilized GPUs. These included six runs on NVIDIA A100-64, three on NVIDIA A100-40, and one on AMD MI100.

\subsubsection{FVOPS per node and energy per iteration}
\begin{figure}[!ht]
  \centering
  \includegraphics[width=0.75\textwidth]{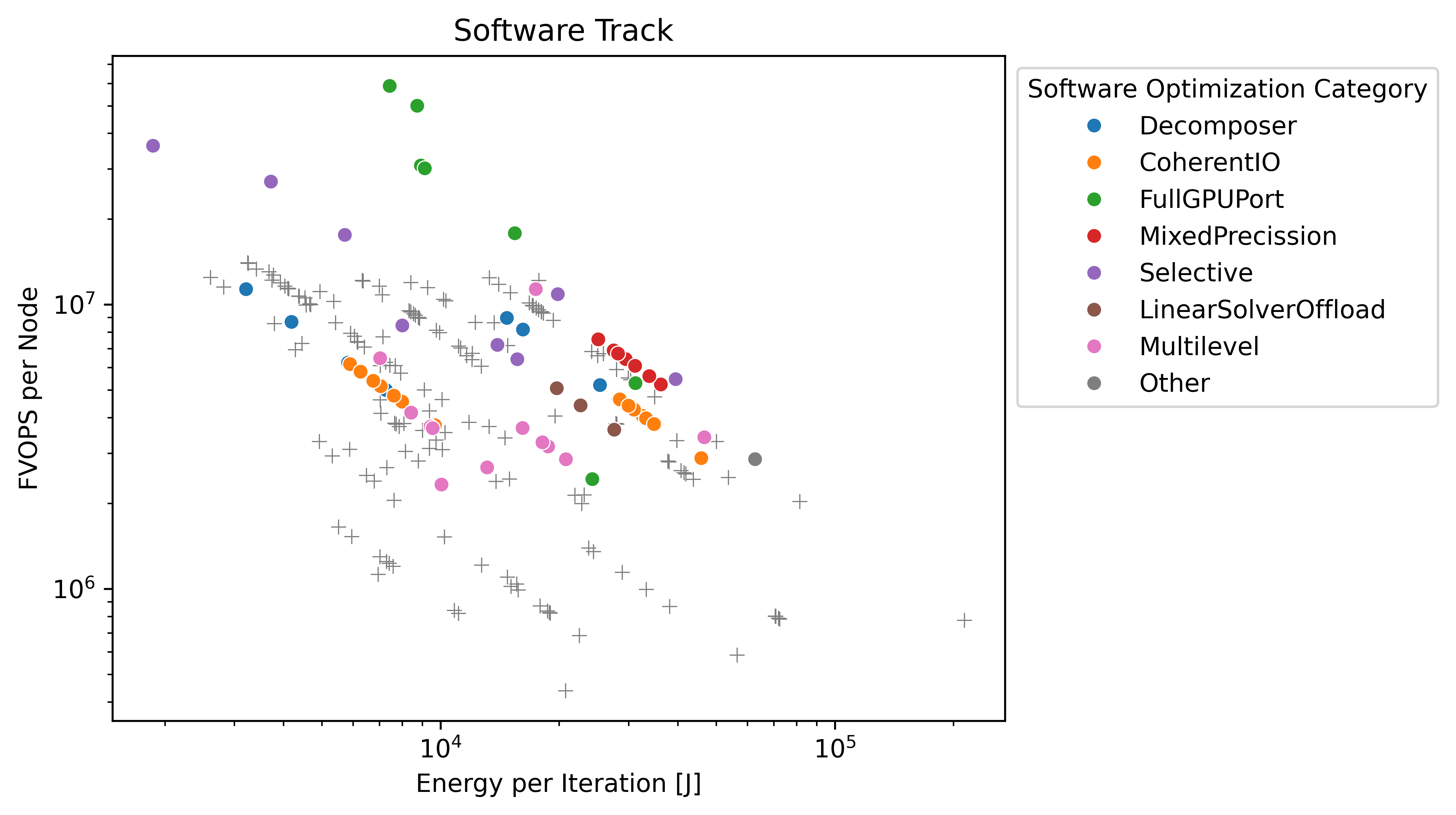}
  \caption{FVOPS per node vs.\ energy per iteration in Joules. Grey: hardware track;
    coloured: software track by optimisation category.}
  \label{fig:overview_hardware_vs_software_track}
\end{figure}

\Cref{fig:overview_hardware_vs_software_track} shows measured FVOPS per node versus
energy per iteration in Joule for all submissions, with  software-track
points coloured by optimisation type and
hardware-track points in grey as reference. Since most submissions investigated strong scaling behaviour, data points tend to cluster along diagonal lines. Several software-track data points can be found in
the upper-left region of the plot, indicating higher throughput per node and a lower energy
consumption per iteration compared to the default \OF version. Here, namely the selective memory allocation optimisation reduced the minimum required electric energy from 2613 Joule to 1867 Joule per iteration. Additionally, full GPU ports and selective-memory optimisations achieved the highest FVOPS per node, with 58 and 36 MFVOPS respectively, with the best software-track result reaching
approximately 4 $\times$ the maximum hardware-track performance of 14 MFVOPS per node.
The increased peak performance is further reflected in the reported minimum wall-clock time required to compute a single timestep. As shown in \cref{tab:minimum_timestep}, the largest improvements are observed for the coarsest mesh, with a reduction of 72\%, followed by the medium mesh, which exhibits a reduction of 44\%.

\begin{table}[!h]
\caption{Reported minimum time per timestep.\label{tab:minimum_timestep}}
\begin{tabular}{l | lll} \toprule
  Mesh & coarse & medium & fine \\    \midrule
Hardware Track    &  0.42  &   0.65     &   0.37   \\
Software Track    &  0.11  &   0.36     &   0.32   \\ \midrule
Change       &   -72 \%   & -44 \%  &  -14\%        \\ \bottomrule
\end{tabular}
\end{table}

\begin{figure}[!ht]
  \centering
  \includegraphics[width=0.75\textwidth]{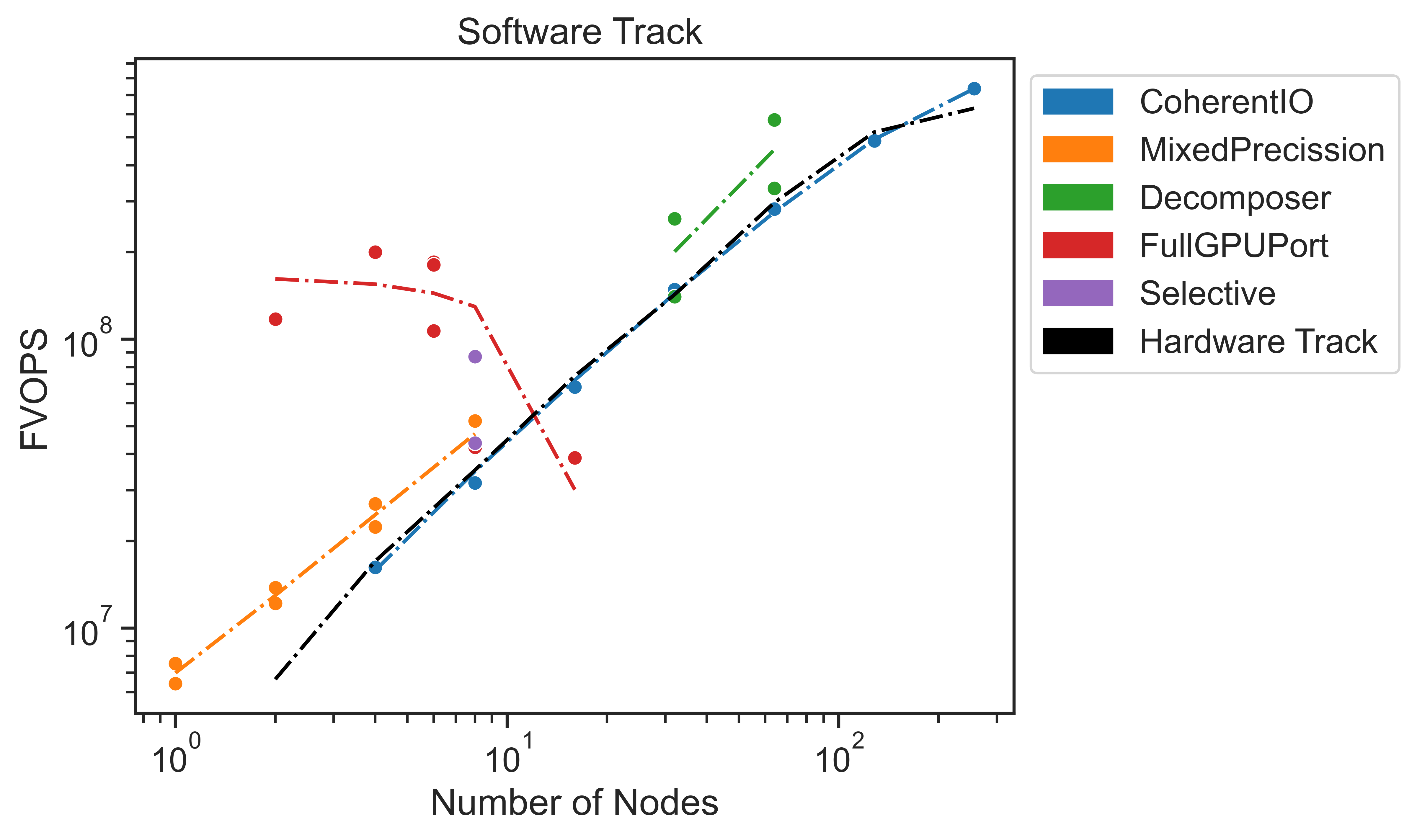}
  \caption{FVOPS vs.\ node count for software-track submissions, coloured by optimisation category. Trendlines using polynomial regression are shown as colored dash-dotted lines. }
  \label{fig:sw_fvops_nodes_gpu}
\end{figure}

It is important to note that, although full GPU ports exhibited the highest per-node performance, their comparatively limited strong-scaling efficiency constrained their overall performance. Consequently, alternative optimisation strategies, such as custom decomposition approaches achieved the highest aggregate performance in terms of FVOPS. This behaviour is illustrated in \cref{fig:sw_fvops_nodes_gpu}, which depicts FVOPS as a function of the number of compute nodes for the fine mesh, categorised by optimisation strategy.
Nevertheless, GPU-based implementations required substantially fewer compute nodes to achieve comparable peak performance relative to CPU-only optimisation approaches, corresponding to a reduction of approximately one order of magnitude (i.e., a factor of 10–15).




\subsubsection{Aerodynamic coefficient consistency}
\begin{figure*}[!ht]
  \centering
  \includegraphics[width=0.75\textwidth]{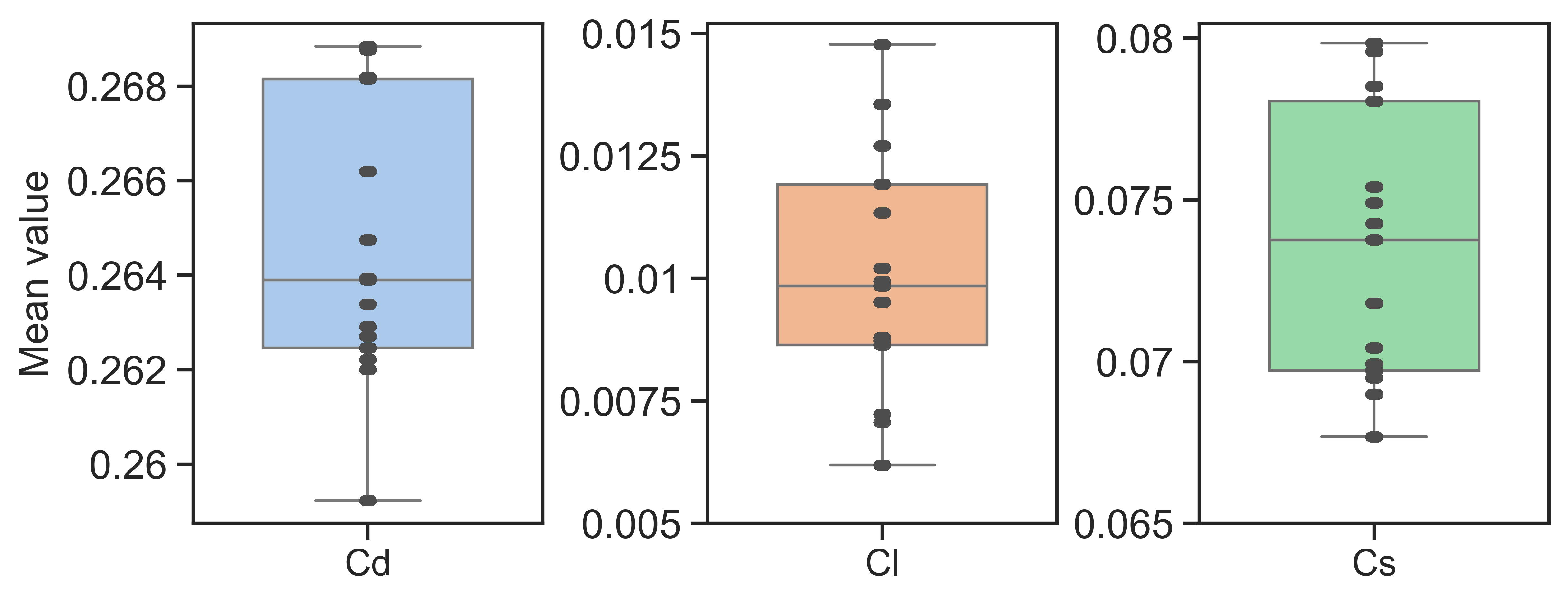}
  \caption{Distribution of mean aerodynamic force coefficients ($C_d$, $C_l$,
    $C_s$) across all submissions.  Box plots show medians and interquartile
    ranges; individual submissions are overlaid as strip points.}
  \label{fig:sw_aero_boxplot}
\end{figure*}

An important quality metric for software-track submissions is the consistency of
the computed aerodynamic force coefficients across different optimisation
approaches and hardware. \Cref{fig:sw_aero_boxplot} shows box-and-strip plots of
the mean drag ($C_d$), lift ($C_l$), and side-force ($C_s$) coefficients across
all submissions. The narrow interquartile
ranges for $C_d$ and $C_s$ indicate that most optimisations preserve solution
accuracy well. The slightly wider spread for $C_l$ reflects the known sensitivity
of lift to solver settings, turbulence-model numerics, and convergence criteria.
Overall, the distributions confirm that software-track submissions maintained
acceptable agreement with the reference values, validating the meanCalc-based
quality checking procedure described in \Cref{sec:validation}.

\section{Conclusion and Outlook} \label{sec:conclusion}

The first OpenFOAM HPC Challenge (OHC-1) collected 237 valid data points from 12
contributing organisations across a hardware track (175 submissions) and a
software track (62 submissions). Participants ran the common occDrivAer
steady-state RANS case on prescribed meshes, with hardware-track runs restricted
to the reference setup and software-track runs allowing approved solver and
decomposition modifications.

\subsection{Hardware-track findings}

The hardware track provided a diverse baseline of \OF performance on current
production systems, covering 25 CPU models from AMD, Intel, and ARM, with runs
from single-node configurations up to 256 nodes. The key findings are:
\begin{itemize}
\item A Pareto front of optimal balance between time-to-solution and
  energy-to-solution was identified, highlighting that increasing parallelism
  yields diminishing time savings at substantially higher energy cost.
\item At small scale, next-generation CPUs (ARM, Intel, AMD) with on-package
  high-bandwidth memory (HBM) and many-core architectures show significantly
  improved single-node performance.
\item At large scale, inter-node communication dominates performance. Many-core
  architectures pose a particular challenge for \OF, which currently relies solely
  on MPI parallelism. A practical strong-scaling limit of approximately
  10\,000~cells per core was observed.
\item The maximum reported hardware-track performance was approximately
  630~MFVOPS per node, with a minimum energy per iteration of about 2613\,J.
\end{itemize}

\subsection{Software-track findings}

The software track demonstrated that substantial gains are achievable through
algorithmic and implementation optimisations. Software-track submissions addressed
multiple optimisation areas, e.g. full GPU ports, GPU linear-solver offloading,
decomposition optimisation, mixed precision, selective memory allocation, and
coherent I/O. Compared to to the hardware track, the computations from the software track achieved:
\begin{itemize}
\item up to 28\% lower minimum energy per iteration (1867 vs.\ 2613\,J);
\item 17\% higher maximum FVOPS (737 vs.\ 629~MFVOPS);
\item up to 72\% faster computation of a single timestep.
\end{itemize}
A key insight is that software optimisations on older-generation hardware can
outperform unoptimised runs on newer-generation CPUs, underscoring the importance
of co-optimising algorithms alongside hardware.

\subsection{Outlook}

OHC-1 establishes a reference dataset and unified metrics (FVOPS, energy per
iteration, time-to-solution) for future comparisons. Contributions exceeded
expectations, but the submission format needs to be tightened for future editions
to further improve comparability. Planned improvements for follow-up challenges
include:
\begin{itemize}
\item encouraging more software-track participation (currently $\sim$20\% of
  total submissions);
\item systematic investigation of network interconnect effects;
\item deriving best-practice guidelines based on the collected submissions;
\item promoting the occDrivAer case as a more relevant community benchmark than
  the commonly used Lid-Driven Cavity;
\item continued publication and curation of the results dataset to enable
  further community analysis (see \Cref{sec:repository}).
\end{itemize}
The HPCTC
repository\footnote{\url{https://develop.openfoam.com/committees/hpc}} remains
the central place for case setups and submission guidelines. All OHC-1
submission data, analysis notebooks, and presentations are publicly available in
the OHC-1 data repository~\cite{ohc1_repo} as described in
\Cref{sec:repository}.


\ifdefined\review
\else
  \section*{Acknowledgements}

\noindent The authors gratefully acknowledge all contributors to OHC-1 for their
submissions and participation in the mini-symposium: Elisabetta~Boella,
Ruggero~Poletto, Eike~Tangermann, Lydia~Schulze, Gabriel~Marcos~Magalh\~aes,
Aleksander~Dubas, Simone~Bn\`a, Stefano~Oliani, and Henrik~Rusche, as well as
the contributing organisations: Wikki~GmbH, University College Dublin, CFD FEA
Service, CINECA, Huawei, Universit\"at der Bundeswehr M\"unchen, Federal
Waterways Engineering and Research Institute, University of Minho, Technical
University of Munich, United Kingdom Atomic Energy Authority, Engys, and E4
Computer Engineering. We also thank the OpenFOAM HPC Technical Committee for
organising the challenge and providing the case setup and infrastructure. Parts of this work were supported by the German Federal Ministry of Education and
Research (grant number 16ME0676K).

  %
  %
  \authorcontributions{
  Conceptualisation, S.L., G.O., and M.W.;
  methodology, S.L., G.O., and M.W.;
  software, S.L., G.O. and M.W.;
  validation, S.L., G.O., and M.W.;
  formal analysis, S.L., G.O., and M.W.;
  investigation,  S.L., G.O., and M.W.;
  data curation,  S.L., G.O. and M.W.;
  writing---original draft preparation, S.L., G.O., and M.W.;
  writing---review and editing, S.L., G.O., and M.W.;
  visualisation, S.L., G.O. and M.W.;
  project administration, S.L., G.O. and M.W.
  All authors have read and agreed to the published version of the manuscript.
  }
\fi


\appendix

\section{Supplementary Material}
\begin{figure}[!ht]
  \centering
  \includegraphics[width=0.85\textwidth]{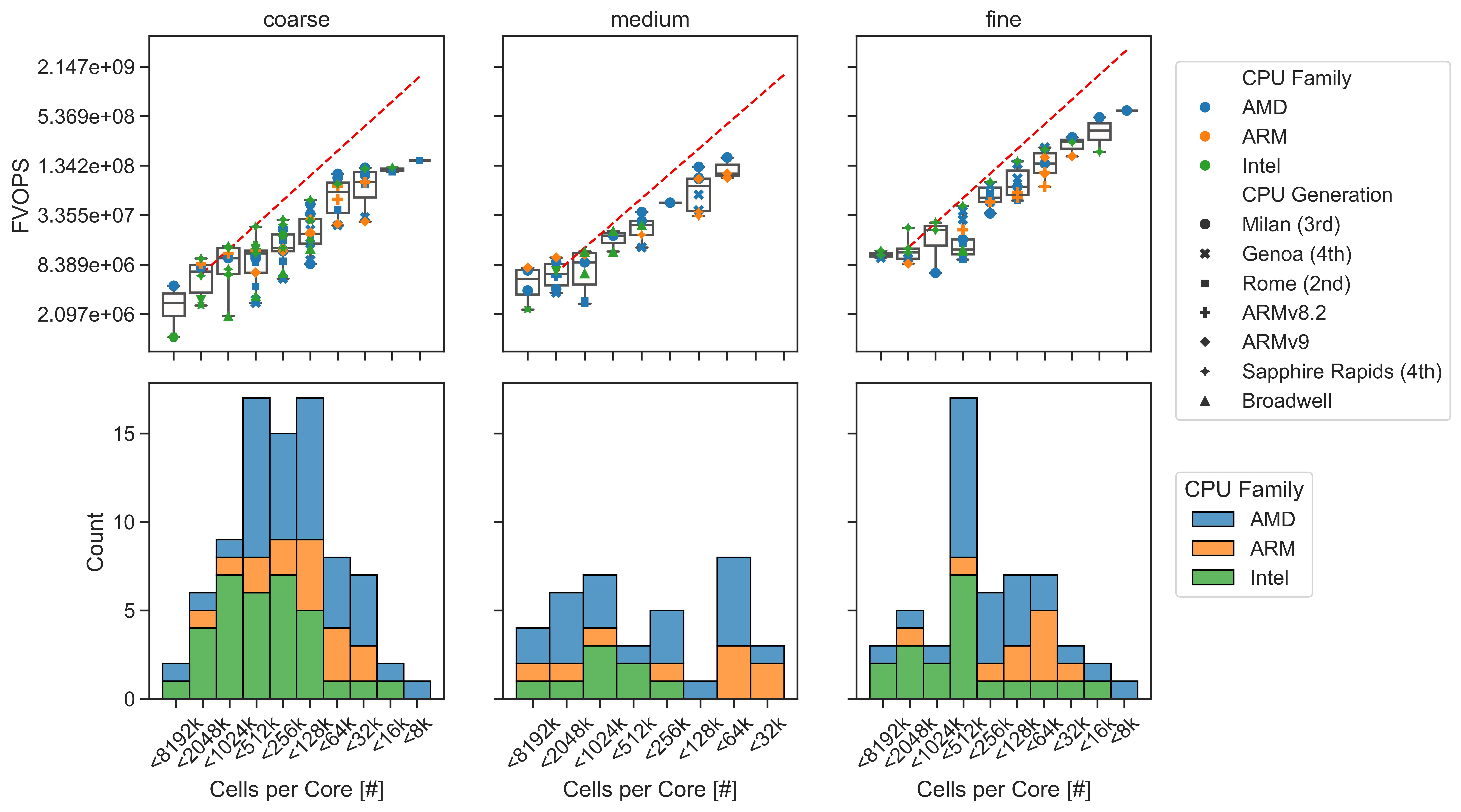}
  \caption{Observed strong scaling behaviour of the FVOPS (top row) over the number of cells per core for the different meshes and the number of data points per bin for each mesh (bottom row).}
  \label{fig:scaling_count_cpu}
\end{figure}

\counterwithin{figure}{section}
An alternative illustration of the strong scaling behaviour shown in \cref{fig:strong_scaling} is provided in \cref{fig:scaling_count_cpu}. In this figure, data points are grouped into bins according to the number of cells per core for each mesh. The top row of \cref{fig:scaling_count_cpu} displays the observed FVOPS, while the bottom row shows the number of data points in each bin. Ideal scaling behaviour is indicated by a red line. Most submissions fall within the range of 100,000 to 1 million cells per core. For both the coarse and fine meshes, only two submissions lie below 16,000 cells per core, and just one falls below 8,000 cells per core. Consequently, the point at which strong scaling begins to deteriorate, i.e. where further reductions in cells per core lead to decreased FVOPS can only be approximated. 
However, the data indicate that FVOPS continues to increase even to 8,000 cells per core.

\FloatBarrier
\section{Data Repository}\label{sec:repository}

While the case setup is available in the HPCTC repository\footnote{https://develop.openfoam.com/committees/hpc/incompressible/simpleFoam/occDrivAerStaticMesh;\\ permalink: \url{https://archive.softwareheritage.org/swh:1:dir:7a4df8317db04b28e1629b10e2a1284782268ec2;origin=https://develop.openfoam.com/committees/hpc;visit=swh:1:snp:f1fb62b1a7c53ee48c5132ea26171a3d7f5f6547;anchor=swh:1:rev:84c262431117f5c921db8335e368a12d0e9fa3f0;path=/incompressible/simpleFoam/occDrivAerStaticMesh/}} all submission data, analysis scripts, and summary presentations from OHC-1 were collected and made
publicly available in a dedicated GitHub
repository~\cite{ohc1_repo}.\footnote{\url{https://github.com/OpenFOAM-HPC-Challenge/OHC1}}
The repository is structured as follows:

\begin{description}
\item[Submissions] The raw submission spreadsheets (Excel files) from all 12
  contributing organisations, together with associated log files, input
  configurations, and force coefficient time series where provided. The
  submission data is licensed under CC~BY~4.0.
\item[Analysis notebooks] A set of Jupyter notebooks used to produce the
  analysis and figures presented in this paper:
  \begin{itemize}
  \item \texttt{Overview.ipynb} --- submission statistics and overview plots;
  \item \texttt{HWTrack.ipynb} --- hardware-track analysis (scaling,
    energy--time trade-off, single-node performance);
  \item \texttt{SWTrack.ipynb} --- software-track analysis (category
    breakdowns, GPU comparisons);
  \item \texttt{IO.ipynb} --- analysis of I/O optimisation submissions;
  \item \texttt{Interactive.ipynb} --- an interactive notebook for custom
    exploration of the full dataset.
  \end{itemize}
\item[Parser utilities] \texttt{OHCParser.py} provides data parsing and
  metric calculation routines that read the raw Excel submissions, compute
  derived quantities (FVOPS, energy per iteration, etc.), and produce a
  unified \texttt{data.json} file for accelerated loading.
\item[Presentations] Participant presentations delivered at the OHC-1
  mini-symposium, as well as summary presentations for both tracks.
\item[Introduction document] A PDF describing the full challenge rules,
  metrics, and data submission guidelines.
\end{description}

The repository enables full reproducibility of the results presented in this
paper and provides a starting point for further community analysis. Researchers
are encouraged to use the interactive notebook or the exported JSON dataset
(\texttt{data.json}) for custom investigations. The source code is released
under the MIT licence. The dataset is archived on Zenodo with DOI
\texttt{10.5281/zenodo.17063427}~\cite{ohc1_repo}.

\bibliographystyle{IEEEtran}

\bibliography{Bibliography}

\end{document}